\documentclass[aps,prl,twocolumn,showpacs,superscriptaddress]{revtex4}
\usepackage{graphicx}

\begin{document}

\title{Dynamics of fullerene coalescence}

\author{Yong-Hyun Kim}
\email[]{yonghyun@mail.kaist.ac.kr}
\affiliation{Department of Physics, Korea Advanced Institute of
Science and Technology, Daejeon 305-701, Korea}

\author{In-Ho Lee}
\affiliation{Korea Research Institute of Standards and Science,
Daejeon~305-600, Korea}

\author{K. J. Chang}
\affiliation{Department of Physics, Korea Advanced Institute of
Science and Technology, Daejeon 305-701, Korea}

\author{Sangsan Lee}
\affiliation{Korea Institute of Science and Technology
Information, Daejeon 305-333, Korea}

\date{\today}

\begin{abstract}

Fullerene coalescence experimentally found in fullerene-embedded
single-wall nanotubes under electron-beam irradiation or heat treatment
is simulated by minimizing the classical action for many atom systems.
The dynamical trajectory for forming a (5,5) C$_{120}$ nanocapsule
from two C$_{60}$ fullerene molecules consists of thermal motions
around potential basins and ten successive Stone-Wales-type bond rotations
after the initial cage-opening process for which energy cost is about 8 eV.
Dynamical paths for forming large-diameter nanocapsules with (10,0), (6,6),
and (12,0) chiral indexes have more bond rotations than 25 with the
transition barriers in a range of 10--12 eV.

\end{abstract}



\maketitle


Carbon nanostructures spanning from fullerenes \cite{Kroto}
to single and multi-wall nanotubes \cite{Iijima91}, nano-peapods \cite{Smith},
and ``Y'' and ``T'' junctions \cite{Terrones02} have attracted much attention
from scientists for last two decades due to their unique structural,
mechanical, and electrical properties \cite{Dresselhaus}.
In particular, carbon nanotubes and related junctions have great potential
for applications on nanometric electronic devices such as quantum functional
transistors and rectifiers since nanotubes can be semiconductors or metals
depending upon their chiralities and diameters \cite{Dekker}.
At the moment, however, development of the nanotube-based technology is limited
because delicate control of their structures is not yet feasible.

Recently, merging processes between carbon nanostructures have been proposed
as a new synthesis technique to construct more complicate functional structures:
(1) Two nanotubes coalesce into diameter-doubled one in a high-energy (1.25 MeV)
electron beam condition \cite{Terrones00}. (2) In the same condition, initially
crossed nanotubes are found to form some functional junctions shaped like
``Y'' and ``T'' characters \cite{Terrones02}.
Since each carbon atom is displaced approximately every 100 sec
by knock-on collisions with the high-energetic electrons, the irradiation-induced
vacancies may play a crucial role in these merging processes.
Terrones and his coworkers \cite{Terrones02,Terrones00} studied the vacancy-related
merging mechanism utilizing the ordinary molecular dynamics technique.
Though introducing various functional structures, this high-energy process may be not
appropriate for nanometric device application because of the uncontrollable defect
formation.
(3) A low energy growth technique for nanotube bundles, where the chiral indexes
of all individual nanotubes are almost perfectly mono-dispersed, was proposed
in the mixture of C$_{60}$ molecules and catalyst Ni pillars with a help
of a $\sim$1.5~T magnetic field \cite{Schlittler}.
The annealing temperature is just 950 $^\circ$C.
(4) In another low energy situation, a single-wall nanotube can be synthesized
by merging fullerene molecules inside a nanotube vessel in
nano-peapod systems \cite{Luzzi,Bandow}.
Of special importance is the last technique of fullerene coalescence
because it is a catalyst-free and bottom-up process, having a potentiality for
diameter and chirality control of single-wall nanotubes.

Nano-peapods \cite{Smith} consist of array of fullerene
molecules (inner peas) and a single-wall carbon nanotube (an outer pod),
and all components are separated from the others by the van der Waals
distance.
The fullerene molecules as an one-dimensional crystal are spaced by
about 0.97 nm or less, and the filling fraction is over 90~\%.
The high fraction is associated with the exothermic encapsulating process
\cite{Okada,Berber}.
Depending on preparation methods, the various fullerene species
can be C$_{60}$, C$_{70}$, C$_{82}$, and metallofullerenes
\cite{Smith,Luzzi,Bandow,Hirahara}.

Electron-beam irradiation \cite{Luzzi} or heat treatment \cite{Bandow}
on these carbon composites induces coalescence of
core fullerenes into an inside single-wall nanotube, of which diameter
is 0.71 nm smaller than that of the outer `vessel' tube.
Since the electron beam in this case is irradiated for
hundreds of seconds and its energy is about 0.1 MeV,
we can not expect the irradiation-induced vacancies
as like Refs. \onlinecite{Terrones02} and \onlinecite{Terrones00}.
Actually, the outer wall was little damaged during the coalescence process
by the electron beams \cite{Luzzi}.
Iijima and his coworkers baked nano-peapods up to 1200 $^\circ$C for
14 hours. They found that the inside fullerenes start to coalesce
at 800 $^\circ$C and complete transforming to a single-wall nanotube
at 1200 $^\circ$C.
They also proposed that diameters of the synthesized nanotubes
are varying from 0.6 to 0.9 nm depending on the size of the outer `vessel'
tubes in a range of 1.3--1.6~nm \cite{Bandow}.
This low energy merging process of fullerenes have great implication
on diameter and chirality control in nanotube growths as we mentioned
above, but little was known about its dynamical process.
Furthermore, the ordinary molecular dynamics study for the process
is not applicable because it never happens within the simulation time
in use of the modern supercomputers.

Instead, topological consideration was made by Ueno {\it et al.}
to the case of a C$_{120}$ capsule with (5,5) chiral index
from two dimerized C$_{60}$ fullerenes \cite{Ueno}.
They found 22 successive Stone-Wales-type bond rotations as a topologically
acceptable pathway for the transformation.
Recently, hypothetic transition states between two consecutive local minima
of the pathway was examined \cite{Noah}.
Similar topological analysis for head-to-head coalescence of
capped nanotubes and fullerenes was made \cite{Zhao}:
(5,5)+(5,5), (10,10)+(10,10), and C$_{60}$+(10,10).
However, these do not guarantee real dynamics of the coalescing process
beyond topology.
Fullerene collision \cite{Zhang,Xia} and nanotube collision \cite{Kawai}
were dealt with molecular dynamics techniques with the assumption of
one constituent's high energetic incidence on the other.
Fullerenes packed well inside single-wall nanotubes (nano-peapods) are
not relevant with such a high energetic incidence.

In this Letter, to simulate the low-energy merging process of
fullerene molecules with appropriate molecular dynamics information,
we use the action-derived molecular dynamics method,
recently proposed by Passerone and Parrinello \cite{Passerone}
with some modifications for many atom systems \cite{YKim}.

The discretized version \cite{Gillilan} of the classical action ($S$)
for numerical simulations is given by
\begin{eqnarray}
S=\sum_{j=0}^{P-1} \Delta (K_j-V_j),
\end{eqnarray}
where $P$ is the total number of steps for discretization of simulation
time $\tau$, $\Delta= \tau/P$ is the constant time interval, and
$K_j$ ($V_j$) are the kinetic (potential) energy of the configuration
${\bf q}_j$ at the $j$-th time step.
The instantaneous velocity is approximated as
${\bf v}_j=({\bf q}_j-{\bf q}_{j+1})/\Delta$, and then $K_j=m{\bf v}_j^2/2$
where $m$ is the mass of each particle.
Dynamical paths can be obtained by finding the stationary point of the
action, i.e., $\delta S$=0, but in general it is a very difficult task
to find the stationary point because it involves a root-finding problem
which requires the second derivative of the potential.
Furthermore, the action is not bounded and the stationary point can be
either a local minimum/maximum, or a saddle.

To obtain effectively a dynamical path that is globally close to the
Verlet trajectory, i.e., an optimal solution for the discretized action,
we employ a rather simple minimization technique, where the action is
supplemented with some penalty functions \cite{Passerone,YKim};
\begin{eqnarray}
\tilde{\Theta} ({\bf q}_j,E;K) = S
+ \mu_E \sum_{j=0}^{P-1} \left( E_j-E \right)^2 ~~~~~~~~~~~~~~~\nonumber \\
+ \mu_{\bf P} \sum_{j=0}^{P-1} {{\bf P}_j}^2
+ \mu_{\bf L} \sum_{j=0}^{P-1} {{\bf L}_j}^2
+ \mu_K \left( \langle K_j \rangle - K \right)^2.
\end{eqnarray}
Here, ${\bf P}_j$ (${\bf L}_j$) is the instantaneous total linear
(angular) momentum, and $E_j=K_j+V_j$ is the instantaneous total energy
of $N$ particles at step index $j$.
$E$ and $K$ is the target total and kinetic energy, respectively,
and all the $\mu$'s are numerical parameters that enforce
the corresponding constraints.
The $\langle\cdots\rangle$ means the time average.
With $\mu_E$, $\mu_{\bf P}$, and $\mu_{\bf L}$, we enforce the conservation
laws of total energy, and total linear/angular momenta on the optimized
trajectory.
The last term in Eq. (2) is associated with the virial theorem that
classical trajectories should satisfy, causing thermal motions around
potential basins; in thermodynamic language,
the virial theorem corresponds to the equipartition theorem.

 \begin{figure}[t]
 \includegraphics[width=8cm]{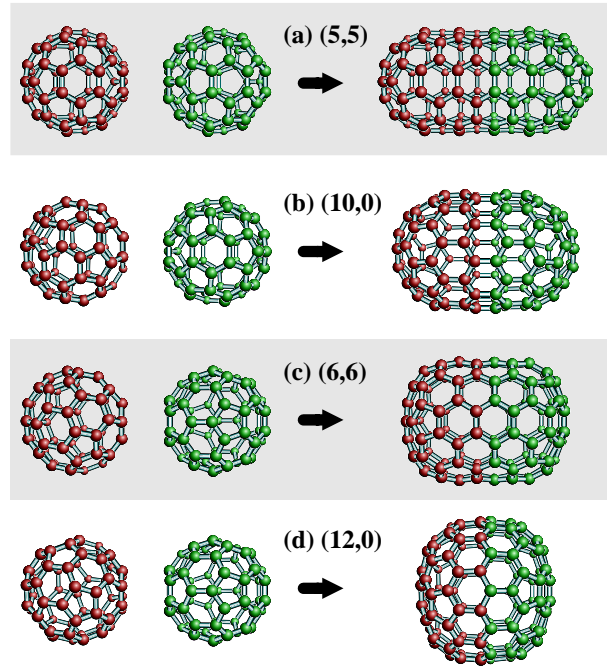}%
 \caption{(color online)
   Initial and final states for forming various C$_{120}$ nanocapsules
   as precursors for fullerene-based nanotube growth.
   \label{fig1}}
 \end{figure}

In our action-derived molecular dynamics simulations, we minimize
the generalized action $\tilde{\Theta}$ with respect to various
atomic paths using a conjugate gradient technique for the given
initial and final $N$-atom configurations, as shown in Fig. 1. The
potential energy and Hellmann-Feynman forces are calculated using
a tight-binding model for carbon systems \cite{Xu}. Details of
formulation and numerical procedure will be presented elsewhere
\cite{YKim}.


For the fullerene coalescing process, we can reasonably assume
that its initial state is two C$_{60}$ molecules separated by
$\sim$1 nm \cite{Luzzi,Bandow}. However, for the final
configuration, we can expect various C$_{120}$ nanocapsules with
different diameters and chiralities because experimentally the
diameters of the synthesized inner tubes ranges in 0.6--0.9 nm
\cite{Bandow}. The simplest case is the C$_{120}$ nanocapsule with
diameter of 0.7 nm and chiral index of (5,5) [Hereafter we call it
the (5,5) nanocapsule], as other workers considered
\cite{Ueno,Noah,Zhao}. We also present possible diameter
thickening process by forming (10,0), (6,6), and (12,0)
nanocapsules with diameters of 0.8--9.8 nm, as shown in Fig. 1.
Assuming that all the atoms move least distance from initial to
final states in order to accelerate the numerical convergence,
we choose the initial orientation of two fullerene molecules
differently depending on the cap structure of the final
nanocapsules.
With this we can remove free rotational motion of a single
molecule in the full molecular dynamics simulation, with saving
much computation time.

 \begin{figure}[t]
 \includegraphics[width=8cm]{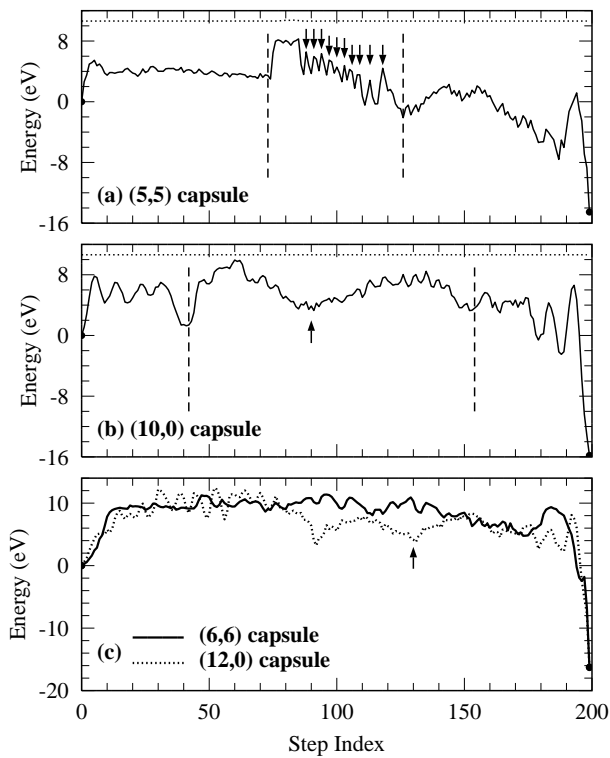}%
 \caption{
   Potential energy variation along the obtained dynamical trajectories
   of fullerene coalescence for forming  various C$_{120}$ nanocapsules.
   In (a) and (b), the solid (dotted) lines indicate the potential
   (total) energy of the (5,5) and (10,0) capsules, respectively,
   along the time step. We divide the trajectories into three parts
   as reactant, activation, and product regions in series.
   In (a), ten small arrows correspond to ten Stone-Wales-type bond
   rotations.
   In (b), the single arrow indicates when the intermediate
   (5,5) capsule arises.
   In (c), we depict the potential energy variation in case of
   the (6,6) and (12,0) C$_{120}$ capsules.
   Also, the small arrow indicates forming the intermediate (6,6)
   capsule in the trajectory of the (12,0) capsule.
   \label{fig2}}
 \end{figure}

We perform action-derived molecular dynamics
simulations using the parameters of $\mu_E$=10$^{8}$,
$\mu_{\bf P}$=$\mu_{\bf L}$=$\mu_K$=$10^4$, $P$=199,
and $\tau$=$P\Delta$=1.2~ps, and display the results in Fig. 2.
With the numerical parameters, we find that the total energy is well
conserved over all time steps for all the nanocapsules, whereas the linear
and angular momenta is well converged for the (5,5) and (10,0) capsules,
but less strictly for (6,6) and (12,0) capsules.
Since $E$ and $K$ cannot be independent variable in a microcanonical problem
where total energy of systems is conserved,
we relax $K$ within the minimization technique,
regarding it as an extra variable after achieving some convergence.

For the coalescing process of the two C$_{60}$ molecules
into the (5,5) C$_{120}$ nanocapsule,
we can divide the obtained dynamical trajectory [See the supplementary movie]
into three parts as
reactant, activation, and product regions in series, as in Fig. 2(a).
In the reactant region around the initial configuration,
the 120 atoms of the two C$_{60}$ molecules are randomly vibrated around their
energy minima with preserving fullerene cage structures.
This motion can be regarded as thermal motion, and be consistent with
random trajectories around potential minima of the Mueller
potential in Ref.~\onlinecite{Passerone}.
The relation $\langle K_j \rangle \approx  \langle \triangle V_j \rangle$,
the virial theorem for the harmonic potential, is satisfied.
In the product region near the final state, we similarly find a thermal
vibrating motion of the (5,5) C$_{120}$ nanocapsule.
The coalescence occurs in-between, i.e., the activation region of about 0.3 ps.
At the transition state, two fullerenes start to merge with breaking some
C--C bonds to open cage, and then a peanut-shaped
intermediate structure appears at step index 87.
Then, we can see 10 successive Stone-Wales-type bond rotations forming
small potential peaks due to the occurrence of undercoordinated carbon atoms.
The ten rotations are attributed to the 5-fold symmetry of the initial
and final states, as discussed in Ref. \onlinecite{Zhao}.
The peak height is about 2--5 eV.
For the whole coalescence process, the energy barrier is estimated to 8.3 eV.
This is quite comparable to the energy barrier for single Stone-Wales
transformation in a C$_{60}$ molecule \cite{Yi}.


For the (10,0) nanocapsule, we also find the reactant and product
regions of the dynamical trajectory with randomly vibrating atomic
motion around both the potential basins. We count about 25
Stone-Wales-type bond rotations in-between activation region,
lasting about 0.6 ps to complete the transformation. An
intermediate (5,5) C$_{120}$ nanocapsule is found at step index 90
on the trajectory for the (10,0) capsule, indicated by the arrow
in Fig. 2(b). Thus, we find a transformation route: two C$_{60}$
fullerenes $\rightarrow$ (5,5) C$_{120}$ nanocapsule $\rightarrow$
(10,0) C$_{120}$ nanocapsule. In the former part, a transition
state occurs with an energy barrier of 9.9 eV, taking about 10
bond rotations, and in the latter part the (5,5) nanocapsule
transforms into (10,0) nanocapsule through about 15 bond
rotations, with an energy barrier of less than 8 eV. Even though
the former process is conceptually same to the coalescence process
for forming (5,5) nanocapsule of Fig. 2(a), the energy barrier is
higher by 1.6 eV because of difference in the initial molecular
orientation as shown in Figs. 1(a) and (b).


In the cases of (6,6) and (12,0) nanocapsules, the whole processes require
more bond rotations than 30.
With the simulation time of 1.2 ps, the obtained dynamical trajectories
shown in Fig. 2(c) entirely belong to the activation process,
different from those for the (5,5) and (10,0) nanocapsules
in Fig. 2(a) and (b).
If we use sufficiently large $\tau$ to cover 30--60 bond rotations,
we may get such thermal vibrating modes near the initial and final states,
but this demands much computational cost.
With the limited simulation time, even though we can not guaranteed the
real dynamics [In fact, total linear/angular momenta are poorly conserved],
we can estimate satisfactorily activation energy barriers within
the action-derived molecular dynamics formalism.
The calculated values are 11.4 (step index 97) and 12.4 eV (step index 50)
for the (6,6) and (12,0) nanocapsules, respectively,
rather high than those for the (5,5) and (10,0) nanocapsules.
The limited simulation time causes the simultaneous Stone-Wales-type
bond rotations, resulting the high energy barriers.
Another reason is that, since we preserve the 6-fold symmetry
in the initial and final states of these systems, double bonds of
the hexagons should be broken to coalesce initially.
This initial opening process needs an energy cost of about 10 eV,
which corresponds to step index 0--30.
We also find an intermediate (6,6) nanocapsule at step index 130
in the path of forming the (12,0) nanocapsule, indicated by the arrow
in Fig. 2(c).

 \begin{table}[t]
 \caption{
   Fullerene coalescence into various C$_{120}$ nanocapsules.
   Diameter of capsules ($D$), formation energy ($H_f$),
   overall potential energy barriers ($V_b$), average kinetic energy
   in Kelvin ($\langle K_j \rangle$), and $O$ values are compared
   for the used target energy ($E$).
   \label{t1}}
 \begin{ruledtabular}
 \begin{tabular}{ccccccccc}
   &  & $D$ ({\AA}) & $H_f$(eV)&$V_b$(eV)& $\langle K_j \rangle$(K) & $O$(a.u.) & $E$ (a.u.) &\\
   \hline
   & (5,5) &  6.85 & -14.53  &  8.3 & 556 & 599.5 & -29.85 &\\
   &(10,0) &  8.02 & -15.81  &  9.9 & 360 & 368.9 & -29.85 &\\
   & (6,6) &  8.32 & -16.25  &  11.4 & 239 & 227.7 & -29.80 &\\
   &(12,0) &  9.78 & -16.30  &  12.4 & 353 & 300.7 & -29.78 &\\
 \end{tabular}
 \end{ruledtabular}
 \end{table}

We summarize all results and the target energy ($E$) used for these
four nanocapsules in Table I.
As the diameters increase, the capsules are more stable with relaxing strain
energy, but estimated energy barriers are higher.
The large energy barriers over 10 eV in the cases of the (6,6) and (12,0)
nanocapsules are attributed to the limited simulation time and
thus simultaneous bond rotations.
Our estimated values for transition barriers are quite reasonable for
low-energy merging process of fullerenes inside single-wall nanotubes.
The value of 10 eV corresponds to system temperature of 1450 K
for two C$_{60}$ molecules, and the experimental temperature for
the merging process is 1500 K \cite{Bandow}.
We also display the average kinetic energy $\langle K_j \rangle$ used for ensuring
the virial theorem, and the discretized Onsager-Machlup actions $O$
\cite{Olender}.
The obtained $O$ values are rather large, especially for
the cases of the (5,5) nanocapsule due to the thermal motions
in dynamical trajectory.
We note that the action-derived molecular dynamics cannot offer
a unique solution for the dynamical paths of a specific rare event.
We may obtain different dynamical paths with various $O$ and $\langle K_j \rangle$,
depending on the target energy, initial seed, and minimization schedule
for the functional Eq. (2).
Nevertheless, we believe that the $\tilde{\Theta}$ trajectory generally reflects
the real dynamics of the system under consideration via the constraints
for the classical laws.
By imposing the appropriate dynamical laws and virial theorem,
we can obtain an estimate, not exact value, for transition
energy barrier and transformation process for specific rare events
based on the real dynamics.


In conclusion, we investigate the molecular transformation of two C$_{60}$
fullerenes into various C$_{120}$ nanocapsules, and find dynamical paths
for the fullerene coalescence using the action-derived molecular dynamics
technique.
In the dynamical process of forming the various C$_{120}$ nanocapsule with
the chiral indexes of (5,5), (10,0), (6,6), and (12,0), we find that
the process consists of successive Stone-Wales-type bond rotations
with the transition energy barrier of about 8--12 eV.
Further growth via coalescence between nanocapsules and C$_{60}$
fullerenes may be natural extension of the above coalescence process with
great similarity, requiring future study.

\begin{acknowledgments}
The authors thank T.-W. Ko. for useful discussion.
This work is supported by the GRID project of KISTI.
\end{acknowledgments}

\end{document}